\documentstyle[onecolumn]{mn}
\newcommand{\be}{\begin{equation}}
\newcommand{\ee}{\end{equation}}
\newcommand{\brhon}{\bar \varrho }
\newcommand{\bv}{\bar v }
\newcommand{\bu}{\bar u }
\newcommand{\bp}{\bar p }
\title[A cosmological hydrodynamic code]
{A cosmological hydrodynamic code based on the Piecewise Parabolic Method}
\author[Claudio Gheller, Ornella Pantano
and Lauro Moscardini]{Claudio Gheller$^{1}$, 
Ornella Pantano$^{2}$
and Lauro Moscardini$^{3}$ \\
$^{1}$ SISSA -- International School for Advanced Studies,
via Beirut 2--4, I--34013 Trieste, Italy\\
$^{2}$ Dipartimento di Fisica {\em Galileo Galilei}, Universit\`{a} di
Padova, via Marzolo 8, I--35131 Padova, Italy\\
$^{3}$ Dipartimento di Astronomia, Universit\`{a} di Padova, vicolo
dell'Osservatorio 5, I--35122 Padova, Italy\\}
\begin{document}

\maketitle

\begin{abstract}
We present a hydrodynamical code for cosmological simulations which uses the
Piecewise Parabolic Method (PPM) to follow the dynamics of gas component and an
N--body Particle--Mesh algorithm for the evolution of collisionless
component. The gravitational interaction between the two components is
regulated by the Poisson equation which is solved by a standard FFT
procedure. 
In order to simulate cosmological flows we have introduced 
several modifications to the original PPM scheme which we describe in detail.
Various tests of the code are presented including adiabatic expansion, 
single and multiple pancake formation and three-dimensional cosmological 
simulations with initial conditions based on the cold dark matter scenario.

\end{abstract}


\section{Introduction}

Most of the present cosmological models are based on the assumption that in the
universe two different kinds of matter are present: the baryonic matter, which
is directly observed and forms all of the bright objects, from stars to the hot
gas present in X--ray clusters, and a dark, collisionless component which
accounts for most of the gravitational mass in the Universe.
The evolution of this system can only be described by treating 
both components at the same time, looking at all of their internal processes and
considering their mutual interaction. In this way, making a suitable choice of 
the
initial conditions, one can describe the evolution of cosmological
structures and estimate all of the related physical observables. This can only
be achieved by using numerical simulations which allow a general
description of the non--linear evolution of the structures. In
particular N--body techniques (Hockney \& Eastwood 1981; Efstathiou et al.
1985; Barnes \& Hut 1986) have proved to be particularly effective
for cosmological problems in which the dynamics is controlled by
gravitational forces as in the case of the dark matter and also for
baryonic structures on very large scales (more than 50$h^{-1}$ Mpc).
With these methods matter is described as a
set of collisionless particles and its dynamics is governed by the
Boltzmann equation. 

The N--body approach, however, is not in general suitable for describing
the behaviour of the baryonic component which is also influenced by
pressure forces, heating and cooling processes. The inclusion of all
these phenomena requires an enormous amount of computational resources as
they act on a very wide range of scales. The thermal input from
gravitationally induced shocks is, for example, maximally effective at
about 10 Mpc, while cooling processes are most important on scales less
than 0.1 Mpc. Therefore to take into account all of these processes
requires a minimum of $10^6$ resolution elements in three dimensions.
The lack of adequate computational resources has delayed the development
of hydrodynamic cosmological codes and only in recent years have a number of
numerical schemes been proposed for following the evolution of
baryonic matter. 

A first family of these techniques derives directly from the N--body
methods. It is called ``Smoothed Particle Hydrodynamics'' (SPH) and represents
fundamental fluid elements in terms of particles. The SPH methods are
intrinsically Lagrangian and, following the fluid elements in their
motion, have high spatial resolution and give an accurate description of
high--density regions, where particles tend to concentrate. On the other
hand they cannot treat properly low--density regions where few particles
are present and mass resolution is poor. For a detailed introduction to
these methods we refer to Hernquist \& Katz (1989), Evrard (1990) and
Steinmetz \& Muller (1993).

Recently Gnedin (1995; see also Gnedin \& Bertschinger 1996) 
has presented a new approach to cosmological 
hydrodynamics called SLH (softened Lagrangian hydrodynamics) which 
utilizes a high resolution Lagrangian hydrodynamics code combined with a 
low resolution Eulerian solver. Its properties are intermediate 
between the Eulerian and SPH approaches.  

Multi--dimensional hydrodynamic codes not based on SPH, 
usually adopt an Eulerian approach and
dynamical equations are solved on a fixed (or adaptive) grid. Mean values of
the fluid quantities are computed in each grid cell by solving the
equations of conservation of matter, momentum and energy density once the
equation of state for the matter is given. Eulerian methods have good mass
resolution, and can describe low--density regions better than SPH
but they are spatially limited by the cell size. An Eulerian
approach has been adopted in the majority of the hydrodynamical 
codes developed for
studying large scale structures (Chiang, Ryu \& Vishniac 1989; Cen 1992; 
Ryu et al. 1993; Bryan et al.
1995; Quilis, Ib\'anez \& Saez 1996; Sornborger et al. 1996). 
For a comprehensive comparison between the Lagrangian and Eulerian approaches
we refer to Kang et al. (1994).

In this paper we describe a cosmological hydrodynamic code that
we have developed using the Piecewise Parabolic Method (PPM) introduced
by Colella \& Woodward (1984). This is a higher order extension of
Godunov's shock capturing method (Godunov 1959; 1961). It is at least
second--order accurate in space 
(up to the fourth--order, in the case of smooth flows and
small timesteps) and second--order accurate in time. The
high accuracy of this method allows minimization of errors due
to the finite size of the cells of the grid and leads to a spatial
resolution close to the nominal one. In a cosmological framework, the
basic PPM technique has to be modified to include the gravitational
interaction and the expansion of the universe. In the gravitational
collapse of cosmological structures, extremely supersonic flows are
generated and so the thermal energy cannot be computed accurately from the total energy, 
as 
is usually done. Thermal energy is only a small
fraction of the total energy and so negative values or a spurious
heating can both be found as a result of numerical errors. This
difficulty has been overcome by computing simultaneously the total and the
internal energy. 

The PPM algorithm has already been used for building a cosmological code
by Bryan et al. (1995), however our approach differs in several respects 
from theirs. In the Bryan et al. work, the one--dimensional time
integration is done by first performing a Lagrangian step and then
remapping the results onto an Eulerian grid expressed in the usual
coordinates comoving with the mean Hubble flow.
We instead adopt a single--step Eulerian scheme.
The construction of the effective left and right states for the Riemann 
problem is then more complicated than in the Lagrangian case, since the 
number of characteristics reaching the edge of a zone is not constant.
On the other hand, this choice allows us to include in the characteristic
equations both the gravitational interaction and the expansion of the 
universe and then the effect of all source terms is accounted for to 
second--order. Moreover, 
an Eulerian approach seems to be preferable in the case of 
complicated three--dimensional structures (Colella \& Woodward 1984).

The hydrodynamical part has been coupled to a Particle Mesh (PM) 
N--body code
that describes the evolution of the dark component. The standard PM code has
been modified to work with a non--constant timestep equal to that used
in the hydrodynamical integration. The coupling is obtained by calculating
the gravitational field due to both the
components by the usual FFT procedure.

The plan of the paper is as follows.
In Section 2 we present the basic equations
written in comoving coordinates. In Section 3 we describe the basic
numerical method and the modifications needed for cosmological calculations.
The numerical tests and results from cosmological simulations are presented in
Section 4, while conclusions are drawn in Section 5. 
A more detailed description of the numerical scheme and of the method 
used for computing
the fluxes are presented in the appendices.

\section{Dynamical Equations} 

The cosmological hydrodynamic equations for a self--gravitating gas in 
coordinates comoving with the expanding universe are:
\be
{\partial \varrho\over \partial t}+{1\over 
a}{\partial\over\partial x_j}\left(\varrho v_j\right) = 0\ ,
\nonumber 
\ee
\be
{\partial\varrho v_i\over\partial t}+{1\over
 a}{\partial\over\partial x_j}
\left(\varrho v_i v_j+ p\delta_{i,j}\right) =
-{\dot a\over a} \varrho v_i-{1\over
 a}\varrho{\partial\phi \over \partial x_i}\ ,\nonumber 
\ee
\be
{\partial E \over \partial t}+{1\over a} 
{\partial \over \partial x_j}[(E+p) v_j] =
-2{\dot a\over a} E-{1\over a} \varrho v_j{\partial \phi
\over\partial x_j}\ ,
%\eqno(EQ3)
\ee
where $x_j$ and $v_j$ are the $j$--th component of the comoving coordinate 
and of the proper peculiar velocity, respectively; $t$ is the
cosmic time; $a$ is the cosmological expansion factor; $\phi$ is the
peculiar gravitational potential; $\varrho$ is the comoving matter density; $E$
is the total energy per unit comoving volume; $p$ is the comoving
pressure.

We treat the baryonic matter as an ideal gas with adiabatic index
$\gamma = 5/3$ and its equation of state is:
\be
p = (\gamma-1)\left(E - {1 \over 2}\varrho {\bf v}^2 \right)\ .
%\eqno(EQ4)
\ee
where ${\bf v}^2$ is the square norm of the velocity vector.

The dynamics of collisionless matter is described in comoving coordinates 
by the following set of equations:
\be
{d x_j\over d t}= {1\over a}{ v_j}\ ,
\ee
\be
{d  v_j\over d t}+{\dot a \over a}  v_j = - {1\over a}{\partial\phi 
\over \partial x_j}\ .
\end{equation}

The peculiar gravitational potential is computed by solving the Poisson
equation: 
\begin{equation}
{\bf\nabla}^2 \phi = {4 \pi G \over
 a}(\varrho_{_{tot}}-\varrho_{_0})\ ,
\end{equation}
where $\varrho_{_{tot}}$ is the total comoving
density (baryonic plus dark matter) and $\varrho_{_0}$ is the comoving mean density.

In the code we have normalized the variables using as basic units the
final time $t_0$, the comoving cell size $x_0$ and the comoving mean
density $\varrho_{_0}$ at time $t_0$. The expansion factor at the final
time is set as $a_0=1$. The other variables are normalized as follows:
$v_j/(x_0/t_0)$, $E/(\varrho_{_0} x_0^2/t_0^2)$, $p/(\varrho_{_0}
x_0^2/t_0^2)$ and $\phi/(x_0^2/t_0^2)$. 
 
The physical temperature $T$ is related to the previous normalized quantities 
by
\be
T = {m_p\over K_B} \left({x_0 \over t_0}\right)^2 {p \over \varrho} \ ,
%\eqno(EQ5)
\ee
where $K_B$ is Boltzmann's constant and $m_p$ is the proton mass. 

\section{Code structure}

\subsection{The PPM Scheme}

The choice of the PPM scheme for the integration of hydrodynamical
equations has been suggested by the particular problems which one has
to face in a cosmological computation. 
The presence of 
highly supersonic flows and the formation
of strong shocks during the collapse of structures require a very accurate
treatment. Moreover the description of phenomena which act simultaneously 
on a very wide
range of scales needs as much resolution as possible with a given number
of cells. 

The PPM algorithm, developed by Colella \&  Woodward (1984),
belongs to a class of schemes in which an accurate representation
of flow discontinuities is made possible by building into the numerical
method the calculation of the propagation and interaction of non--linear
waves. There are two possible formulations of the PPM scheme:
hydrodynamical equations can be solved either in a single Eulerian step
or with a Lagrangian step followed by a remapping  onto the Eulerian
grid. We have adopted the first approach as it appears more suitable 
for problems with complicated spatially dependent source terms.
A detailed presentation of the finite difference equations 
used in the code are presented in Appendices A1 and A2: here we describe
the basic operations characterizing the PPM scheme. 

In the PPM
method we start from the knowledge of a set of zone--averaged values of
hydrodynamical quantities. To update these averages we solve
hydrodynamical conservation equations: this requires the estimate of the
fluxes of the conserved quantities at the zone interfaces which is done
according to the following one--dimensional procedure. First, we
construct a piecewise parabolic (third--order) one--dimensional
interpolation function for each of the hydrodynamical variables and for the
source terms. This leads to a more accurate representation of smooth
spatial gradients as well as a sharper description of discontinuities.
The interpolation polynomial $w$ is computed such that: 
\be
w_j\ =\ {1\over\Delta x}\int_{x_{j-1/2}}^{x_{j+1/2}}w(\xi)d\xi\ ,%\eqno(P5)
\ee
where $w_j$ is the mean value of the interpolation polynomial over the
$j$--th zone, $x_{j\pm 1/2}$ are the positions of the zone interfaces and
$\Delta x=x_{j+1/2}-x_{j-1/2}$. It is also required that inside each
zone $w(\xi)$ is continuous and monotonic. 

The interpolated distributions can then be used to estimate properly the
average values of the variables to the left and right of each zone edge.
This is performed in two steps. In the first step, one determines the
characteristic domain of dependence with respect to a zone edge, that is
the regions of space to the left and to the right of the edge that
contain all of the information that can reach the zone edge during the
current timestep. Then, using the interpolated distributions, the mean
values of hydrodynamic variables are computed over the domains of
dependence on each side of the edge. In the second step, we solve the
Riemann problem using as initial left and right states the mean values
of hydrodynamical variables computed previously on each side of the zone
edge. The solution of this gives the time averaged values of the variables at
the zone edge which are then used to compute inter--cell fluxes and
hence the fluid variables at the new time. This closes the basic
one--dimensional PPM loop.

The extension to more dimensions is achieved using a direction splitting
procedure (Strang 1968). If we indicate with $L_k$ the integration in
the $k$--th direction, the updated value for the generic hydrodynamical
variable $u$ is 
\be
{u}^{new}\ =\ L_z L_y L_x{u}^{old}\ .
%\eqno(O1)
\ee
In order to preserve second--order accuracy in space, in successive timestep
the order of directional integration is permuted as follows:
\be
L_z L_y L_x,\ L_xL_yL_z,\ L_xL_zL_y,\ L_yL_zL_x,\ L_yL_xL_z,\ 
L_zL_xL_y\ . 
\ee
For cosmological applications several important modifications must be
introduced into the basic PPM scheme. The inclusion of gravitational forces
and cosmic expansion changes the form of the usual hydrodynamical
equations. Expansion of the universe enters the hydrodynamical equations
in two different ways. First, all of the fluxes and the source terms are
multiplied by the factor $1/a(t)$. Second, two further terms appear:
$(\dot a/a)\varrho v_j$ and $(\dot a/a)E$ in the equations for
conservation of momentum and energy, respectively (in the following we
will refer to these terms as expansion terms). 
This leads to important changes both in the
Riemann solver, as we will show in the next section, 
and in the final integration.
In this last step, forces and expansion terms
are treated in different ways. First hydrodynamical
quantities are updated
without considering the expansion terms. These in fact, being
proportional to the integrated quantity itself, cannot be included in
the basic integration procedure. The time centered values of gravitational 
forces are
calculated by finite differencing the potential field computed by linear
extrapolation at $t^{n+1/2}$. In
order to obtain a proper estimate of the force felt by a fluid element, its
averaged value over the entire zone is calculated. Finally, 
we compute the effect of expansion terms by using a semi--implicit 
scheme (see equations A8-A10 and A17).

\subsection{Characteristic Equations}

The solution of the Riemann problem, which is the central feature 
of the Godunov approach, requires knowledge of the characteristic
form of hydrodynamic equations.  Because of our choice of coordinates 
comoving with the expanding universe,
the finite difference expressions which we use in the Riemann 
solver differ slightly from those discussed in Woodward \& Colella 
(1984). In this section we write the characteristic equations for the 
system (1)--(3), while in Appendix A2 we present the scheme used in the 
code for computing hydrodynamic fluxes. 

For the time integration we use a directional splitting procedure. 
Indicating by $x$
the relevant space direction during the one--dimensional
sweep, by $v$ the velocity in the direction of the one--dimensional
sweep and by $u$ the velocity orthogonal to $v$, 
equations (1)--(3) can be rewritten in the following
non--conservative form: 
\be
{\partial{\bf U}\over \partial t} + {\cal A}{\partial{\bf U}\over \partial x}
+ {\bf C}\ =\ 0\ ,
\ee
where
\be
{\bf U}\ =\ \left(\matrix{\varrho\cr v\cr u\cr p\cr}\right)\ ,\qquad
{\cal A}\ =\ {1\over a}\left(\matrix{v&0&0&\varrho\cr
0&v&0&1/\varrho\cr 
0&0&v&0\cr 
0&\gamma p&0&v\cr}\right)\ ,\qquad
{\bf C}\ =\ \left(\matrix{0\cr 
v\dot a/a-g\cr
u\dot a/a\cr
2p\dot a/a\cr}\right)\ .
\ee
In order to find the characteristic form of this system, we 
first solve the corresponding eigenvalue equation:
\be
{\rm det}\left ({\cal A} - \lambda {\cal I}\right)\ =\ 0
\ee
obtaining as eigenvalues: 
\be
\lambda_0\ =\ {v\over a}\ ,\qquad \lambda_-\ =\ {v-c\over a}\ ,\qquad
\lambda_+\ =\ {v+c\over a}\ ,
\ee
where $\lambda_0$  is a double solution and $c$ is the sound speed.
The eigenvalues are used to compute the domain of dependence of a given fluid
element. Note that in expanding coordinates the domain must be rescaled 
by the
factor $a(t)$ with respect to that for a non--expanding system.

The corresponding 
left eigenvectors ${\bf l}_k$ are:
\be
{\bf l}_{01}\ =\ (0,0,1,0)
\ee
\be
{\bf l}_{02}\ =\ (1,0,0,-1/c^2)
\ee
\be
{\bf l}_-\ =\ (0,-{\varrho c \over 2},0,{1 \over 2})
\ee
\be
{\bf l}_+\ =\ (0,{\varrho c \over 2},0,{1 \over 2})\ .
\ee
The characteristic form of hydrodynamical equations is then:
\be
{\bf l}_k{\partial {\bf U}\over \partial t}+{\bf l}_k{\cal A}{\partial 
{\bf U}\over \partial x}+{\bf l}_k{\bf C} \ dt\ =\ 0\ ,
\ee
\be
{\bf l}_k{\partial {\bf U}\over \partial t}+\lambda_k{\bf l}_k{\partial 
{\bf U}\over \partial x}+{\bf l}_k{\bf C} \ dt\ =\ 0\ ,
\ee
that is:
\be
{\bf l}_kd{\bf U} + {\bf l}_k{\bf C} \ dt\ =\ 0,\qquad {\rm along}\ 
{dx\over dt} = \lambda_k\ .
\ee
For our fluid, the characteristic equations are:
\be
c^2 d\varrho-dp-2{\dot a\over a}p \ dt\ =\ 0
\ee
\be
du -{\dot a\over a}u\ dt \ =\ 0
\ee
along ${dx/dt} ={v/a}$, and
\be
dp \pm \varrho c~ dv+ \left[ 2{\dot a\over a}p \pm 
\varrho c \left({\dot a\over a}v- g\right)\right]dt\ =0 
\ee
along ${dx/ dt}=(v \pm c)/ a$.

Characteristic equations are used both for calculating the 
domain of dependence of a zone edge and in the construction of 
the Riemann solver where we take account, to second--order, of the gravity and
expansion (see Appendix A2). The solution of the Riemann problem gives the
required time--averaged values of the hydrodynamical variables which 
are used in the final integration step. 

\subsection{Correction to the Energy}

In order for overall energy to be accurately conserved it is important
to use the total energy density $E$ directly as a dependent variable
(see equation 3) rather than calculate it as the sum of the kinetic 
and internal energy densities (${1\over 2}\varrho{\bf v}^2$ and $e$,
respectively). The value of $e$ is however required in order calculate
the thermodynamic quantities such as pressure and temperature and the original 
PPM code calculates $e$ as the difference between $E$ and ${1\over 2}
\varrho {\bf v}^2$. However, there is a problem with this for cosmological
simulations where highly supersonic flows are often present. In these 
situations, the ratio between the kinetic and internal energies
can reach values up to $10^8$ in which case the errors in calculating $e$
by the standard PPM procedure become much larger than the quantity itself.
 
In our code, we solved this problem by calculating both the total and 
internal energy per unit comoving volume.
In comoving coordinates the equation for the internal energy
density $e=p/(\gamma-1)$ is the following:
\be
{\partial e\over\partial t}\ =\ -{1\over a}{\partial\over\partial x_k}
(ev_k)-2{\dot a\over a}e-{1\over a}p{\partial v_k\over\partial x_k}\ .
%\eqno(E1)
\ee
Hence, at each
timestep, both the total and internal energy equations are solved and the
results combined according to the local conditions in the 
fluid.
The value obtained from equation (26) has to be only used 
when  the internal energy cannot be calculated correctly as 
the difference between total and kinetic energy. This is usually the case 
in expanding or weakly compressing (comoving) regions,
that is in regions in which one of the following conditions holds:
\be
{\bf \nabla \cdot v}\ \ge\ 0\ ,\qquad
{1\over p}\vert{\bf\nabla} p\vert \le\ {\eta_1}\ .
\ee
If neither of these conditions is verified, the internal energy is computed
according to the following criterion: 
\be
e =\cases{
E-{1\over 2}\varrho {\bf v}^2\quad 
& if $E-{1\over 2}\varrho{\bf v}^2/ E
\geq \eta_2$\cr
e & otherwise\ ,\cr}%\eqno(E5)
\ee
where $\eta_1$ and $\eta_2$ are suitable parameters smaller than unity, fixed 
on the bases of numerical experimentation.
Various tests have suggested the use of the following values:
$\eta_1=0.3$ and $\eta_2=0.005$. 

\subsection{N--body Code}
 
The dynamics of the dark component is described using a PM N--body
code (Hockney \& Eastwood 1981). In order to solve equations (5)--(6) the 
standard
second--order leapfrog method has been replaced by a second--order two--step
Lax--Wendroff method (see Appendix A1). The leapfrog scheme requires a
constant timestep, while hydrodynamics makes it desirable to have 
a variable time interval. Density
and forces are calculated by 
using the cloud--in--cell interpolation scheme.

\subsection{Timestep}

In determining the new timestep at the end of each loop we impose 
several constraints. First of all, we require that the maximum variation 
of the expansion factor $a$ during a timestep is less than $2\%$. 
We impose also that dark matter particles move no more than half a
cell in a single step and that the baryonic dynamics satisfies
the Courant condition:
\be
\Delta t\ \le\ \min\left [{C_c \ a(t)\Delta x\over [c+\max(\vert v_x\vert,
\vert v_y\vert,\vert v_z\vert)]}\right ]\ ,%\eqno(T2)
\ee
where $C_c$ is the Courant number and the minimum is computed over all
cells. We set $C_c=0.5$. 
Finally, we require that in a single step the 
timestep increases by no more than $20\%$ from its previous value. 

Our tests indicate that the expansion condition is important only in the 
first part of the simulation, when the amplitude of the fluctuations 
is still small. At later times the Courant condition becomes the dominant
one.

\section{Tests of the code}

Our code has been subjected to a series of numerical tests, ranging 
from the purely
hydrodynamical to cosmological ones, in one and more dimensions. In each
case conservation of energy has been checked. For the cosmological tests this
check has been performed by using the cosmic energy equation which
for an ideal gas in an expanding frame is (Peebles 1980):
\be
{d\over dt}(E+W)+{\dot a\over a}(2E+W)\ =\ 0\ ,%\eqno(T1)
\ee
where $E$ is the total (internal plus kinetic) comoving energy and $W$
is the gravitational energy of the gas.

Equation (30) can be rewritten as:
\be
{d\over dt}[a(E+W)]+\dot a E\ =\ 0%\eqno(T2)
\ee
and integrated with respect to the expansion factor giving:
\be
a_2E(a_2)-a_1E(a_1)+\int_{a_1}^{a_2}E\ da\ =\ 
-[a_2W(a_2)-a_1W(a_1)]\ ,%\eqno(T3)
\ee
where $a_1$ and $a_2$ are the values of the expansion factor at two
different times.

We checked energy conservation computing the quantity:
\be
R\ =\ {a_2E(a_2)-a_1E(a_1)+\int_{a_1}^{a_2}E\ da\over 
-[a_2W(a_2)-a_1W(a_1)]}\ .%\eqno(T4)
\ee
For perfect energy conservation R should remain as unity. The results obtained
in our tests are presented in the following subsections.

\subsection{One--Dimensional Shock Tube Test}

In Fig. 1 we present the results of a standard shock tube test. In
this case we solve the hydrodynamical equations (1)--(3) without the
source terms on the right hand side and with $a=1$. Initially the gas is
at rest ($v=0$) and density and pressure show a central jump from
$\varrho_{_l}=1$, $p_l=1$ to $\varrho_{_r}=0.1$, $p_r=0.1$. The calculation is
performed using a grid of 128 zones, the initial time is $t_i=1$ and the
final time, shown in the figure, is $t_0=20$. 

The numerical results (open circles) are in very good agreement with the
analytical ones (solid lines). In particular, shocks are typically resolved
within one cell and contact discontinuities are represented in four or five
zones. In order to reduce the diffusion of contacts we have used, for the
density field, the discontinuity detection algorithm of Colella \& Woodward 
(1984). We have not introduced artificial diffusion and/or numerical
flattening since numerical ripples around steep gradients
are either absent or very small and do not affect the solution during the whole
evolution. 

\subsection{Adiabatic Expansion}

After having tested the hydrodynamic part of the code alone, we have introduced
expansion and gravity and have concentrated on tests describing physical
situations that are expected in cosmological simulations. In all of 
the following
tests we have assumed an Einstein--de Sitter pure baryonic universe with
vanishing cosmological constant and we fixed the value $H_0=50~$km s$^{-1}$
Mpc$^{-1}$ for the Hubble constant. The comoving
size of the computational box is $L=64~h^{-1}$Mpc. For this model our basic
units of normalization are:
\be
\varrho_{_0}\ = \ {3 H_0^2\over 8\pi G} 
\qquad 
t_0\ =\ {1\over (6\pi G\varrho_{_0})^{1/2}}\ ,
\ee
where $G$ is the 
gravitational constant. The expansion factor depends on the
cosmological time as 
\be
a(t)\propto t^{2/3}\ .
\ee

As a first test we have simulated the adiabatic expansion of an unperturbed
universe, starting at time $t_i=0.01$ 
from a homogenous distribution of matter
with velocity $v_i=100$ km s$^{-1}$ 
and temperature $T_i=4.6$ K. In this situation
equations (2)--(3) are easily integrated giving: 
\be
v\ =\ v_i(a_i/a)\ ,\qquad T\ =\ T_i(a_i/a)^2\ ,%\eqno(T5)
\ee
where $a$, $v$ and $T$ are the
expansion factor, the velocity and the temperature respectively, at time $t$.
At the final time $t_0=1$ numerical and analytical solutions are compared
and the errors calculated as:
\be
\Delta T\ =\ {\vert T_{num}-T_{an}\vert\over T_{an}}\qquad 
\Delta v\ =\ {\vert v_{num}-v_{an}\vert\over v_{an}}\ ,%\eqno(T6)
\ee
where the subscripts $num$ and $an$ indicate, respectively, 
the numerical and
analytical values of the variables.
We have indicated with $\Delta a$ the maximum fractional variation of 
the expansion factor in a timestep. 
Using $\Delta a=0.1$ 
we have obtained $\Delta
T=6\times 10^{-3}$ and $\Delta v=3\times 10^{-3}$. The errors decrease to
$\Delta T=8\times 10^{-5}$ and $\Delta v=4\times 10^{-5}$ for $\Delta a=0.01$. 
The value of the parameter $\Delta a$ used in the simulations is fixed 
by considering the balance between accuracy and computational time.

\subsection{Zel'dovich Pancake}

In this test we describe the formation of a one--dimensional pancake. The test
is particularly important for cosmological studies as all of the related physics
(hydrodynamics, expansion, gravity) is included. 
In addition, it represents the evolution of an initial sinusoidal wave 
and so can be seen as a single--mode analysis of the full 
three--dimensional problem.

At the initial time, corresponding to $a_i=0.01$, 
a sinusoidal velocity field has been imposed on the computational grid.
The amplitude of the initial
perturbation has been chosen such that the caustic forms at $a_c=0.5$. At $a_i$
the density and temperature fields are uniform. We set $\varrho=1$ and 
$T=100$ K. 
The final time corresponds to
$a_0=1$. The same test has been repeated using eight different numbers of zones
ranging from $N_{c}=8$ to $N_c=1024$.

During the first part of the evolution, 
we are in the linear phase and the numerical
results can be directly compared with the analytical 
solution which applies for a non--collisional fluid since the effect 
of the pressure is 
negligible in this phase. This comparison gives an estimate of
the accuracy of the code. 
Moreover, using results obtained 
with different numbers of zones we can study the convergence of the numerical
solution.
The error is computed as:
\be
\Delta\varrho\ =\ {1\over N_g}\sum_{i=1}^{N_g} {\vert 
\varrho(x_i)-\varrho_e(x_i)
\vert\over\varrho_e(x_i)}\ ,%\eqno(T7)
\ee
where $\varrho_e$ is the exact solution. 
We found that at $a\sim 0.1$,
when shocks are not present, 
the error is below 1$\%$ already with only 16 
zones.

The non--linear phase of the evolution is the most interesting. The strong
shocks and large gradients that form are in fact a severe test for any
Eulerian numerical code. The results at $a_0=1$ obtained 
using grids of 256 and 1024
zones are compared in Fig. 2. The only important difference between the two
solutions is the height of the central density peak, but this was expected
as a consequence of the different resolutions in the two cases. All of the other
features the solution are in very good agreement between the two 
simulations showing the convergence of the numerical solution. 
Shocks are resolved in one zone and there are no numerical postshock ripples.

In Fig. 3 we present the results of the energy conservation test. We have
calculated at the final time 
the conservation parameter $R$ (defined in equation 33) and the total
gravitational ($W$), kinetic ($E_{KIN}$) and thermal ($E_{TH}$) 
energy of the fluid when a various
number of cells $N_c$ are adopted. The energy is plotted in the code units.
The error in the energy conservation is of about 5$\%$ for $N_c=32$
and decreases to $0.1\%$ for $N_c=1024$. The behaviour of the three energy terms
shows that at least 32 zones are needed for starting 
to see the convergence of the
solution. The poor resolution of the 8 and 16 zone cases, in fact, makes it
impossible to describe correctly the narrow central peak, which 
remains too diffused. The shallower central potential wells induce a 
smaller acceleration in the infalling matter. The compression is low and
the transformation from kinetic to internal energy is inefficient and so 
produces a pressure which is too low to support the infall of the matter.
However the collapse of the peak is prevented by the artificial effect
of the coarseness of the grid. 
 
\subsection{Multiple Pancake Formation.}

The last one--dimensional test which we have performed consists of simulating 
the evolution of a random initial density field.
In this test different wavelengths evolve together; hence we can verify
the ability of the code for describing shock interactions and the merging 
of small structures. 

The initial conditions are set by perturbing the density field according
to a Gaussian random distribution characterized by a power--law 
spectrum of the general form
\be
P(k) = P_0 k^n {\rm exp} (-k^2 R_f^2) \ ,
\ee
where $P_0$ is the normalization constant and $n$ is the spectral index. The
short wavelength cut--off at the scale $R_f=2$ grid--points ensures that the
results are not affected by the sampling of modes whose size is close to that
of the resolution of the simulation. 

The normalization constant is fixed requiring that the one--dimensional
variance $\sigma^2$ at the final time $t_0$ (corresponding to the expansion
factor $a_0=1$) for linearly evolved perturbations 
is equal to unity on the scale
$R_*=10^{-1}L$ so that: 
\be
\sigma^2(R_*)\ =\ {1\over\pi}\int_0^{\infty}P(k)W(kR_*)\ dk\ =\ 1\ ,
%\eqno (MP2)
\ee
where the function $W(kR_*)$ is a one--dimensional top--hat filter:
\be
W(kR_*)\ =\ {\sin (kR_*)\over kR_*}\ .%\eqno(MP3)
\ee
The initial time $t_{i}$ is chosen by imposing that, at $t_{i}$, the largest
density fluctuation is equal to unity, so that the system can be considered
still evolving linearly.

As an example, we show in Fig. 4 the results at the final time $t_0$ 
for a simulation 
with spectral index $n=1$ when $N_c=256$ is used, i.e. with the same number of
cells used in the simulation of the single pancake. 
Shocked regions, characterized by the presence of typical double peaked maxima
in the temperature field and by the smoothing of steep negative gradients in the
velocity field, are at temperatures ranging from $10^4$ to $10^6$K.
Underdense regions are instead at much lower temperature. Collapsing regions
surrounding density peaks are, at the beginning, at low temperature, but heat
up as soon as propagating shocks reach them. The errors in energy 
conservation are
around $3\%$, slightly higher than the value obtained for the Zel'dovich
pancake test. 

The different clustering properties of baryonic and dark matter and the
multiple pancake formation have been studied in much more detail in Gheller,
Moscardini \& Pantano (1996) by using numerical simulations based on a
one--dimensional version of the PPM+PM code presented here. In that paper
power--law initial power spectra with $-1 \le n \le 3$ have been simulated using
8192 zones and 16384 particles. The results show that at large scales the final
distribution of the two components is very similar while at small scales the
dark matter presents a lumpiness which is not found in the baryonic matter. 

\subsection{Cold Dark Matter Universe}

As first checks of our complete fully three--dimensional code, 
we repeated the tests
of the one--dimensional shock tube and of the Zel'dovich pancake formation.
Using a $64^3$ grid, the results (not shown here) are consistent with those
obtained with a similar resolution in one--dimension. Finally we have simulated
the evolution of a purely baryonic Einstein--de Sitter Universe with an initial
cold dark matter (CDM) power spectrum. As in the previous tests, we fixed the
Hubble constant to be $H_0=50~$km s$^{-1}$ Mpc$^{-1}$ and the comoving size of
the computational box to be 
$L=64~h^{-1}$Mpc. In this way our numerical results can
be compared with those obtained by  Kang et al. (1994), even if a different
random realization of the initial conditions has been used. Perturbations are
initially Gaussian distributed and the primordial density power spectrum can be
written in the form: 
\be
P(k)\ =\ P_0 k^n T^2(k)\ ,
\ee
where $T(k)$ is the CDM transfer function given by the 
Davis et al. (1985) fitting formula:
\be
T(k)\ =\ (1+6.8k+72.0k^{3/2}+16.0k^2)^{-1}\ .
\ee
In the standard CDM model,
the primordial spectral index has the Zel'dovich value $n=1$. 

The normalization constant $P_0$ is fixed
such that the linear mass variance at the present time is unity 
in a sharp--edged
sphere of radius $R_8=8 h^{-1}$ Mpc:
\be
\sigma^2(R_8)\ =\ {P_0\over 2\pi^2}\int_0^{\infty} k^{n+2}T^2(k)
W^2_{TH}(kR_8)~ dk=1\ .
\ee
In the above equation, $W_{TH}(kR)=(3/kR)j_l(kR)$ 
is a top--hat window function and $j_l$
denotes the $l$th--order spherical Bessel function. 
The velocity field at the initial time $t_i$ 
has been calculated in the linear perturbation
approximation (Peebles 1980) as:
\be
{\bf v} = {2\ {\bf\nabla}\Phi\over 3 H(t_i) a(t_i)} \ .
\ee

We are interested in checking the convergence properties of  our code, so we
evolved the same initial conditions on three different grids with $32^3$,
$64^3$ and $128^3$ cells, respectively. 
The initial conditions are set up according to the previous prescriptions on
the coarsest grid. The initial time, fixed by the condition that the maximum
density fluctuation is unity, corresponds to a redshift $z\sim 20$. 
We run our simulations on a IBM RISC 6000/590.  

\begin{table}[tp]
\centering
\caption[]{
The simulations. Column 1: the number of zones in each dimension $N_c$;
Column 2: the mean temperature $\langle T\rangle$. Column 3:
the average
mean temperature $\langle T\rangle_{\varrho}$.
Column 4: the average temperature weighted by the
density squared $\langle T\rangle_{\varrho^2}$.
Column 5: the r.m.s. of the density
fluctuation field $\sigma$.
} 
\tabcolsep 5pt
\begin{tabular}{lccccc} \\ \\ \hline \hline
 $N_c$ & $\langle T\rangle$ & $\langle T\rangle_\varrho$ & $\langle 
T\rangle_{\varrho^2}$ & $\sigma$ 
\\ \hline
32 & 1.48 & 6.02 & 19.49 & 2.17 \\
64 & 2.02 & 10.61 & 28.63 & 2.48\\
128 & 2.02 & 11.91 & 32.95 & 2.77 \\ \hline

\end{tabular}
\end{table}

We have first analyzed some integral properties of the results at the final
time. We have calculated the mean temperature $\langle T\rangle$, the average
mean temperature $\langle T\rangle_{\varrho}\equiv \langle
T\varrho\rangle/\langle\varrho\rangle$, the average temperature weighted by the
density squared $\langle T\rangle_{\varrho^2}\equiv \langle
T\varrho^2\rangle/\langle\varrho^2\rangle$ and the variance of the density
fluctuation field $\sigma^2\equiv (\langle \varrho^2 \rangle /
\langle\varrho\rangle^2 - 1)$. As in Kang et al. (1994), the calculation has
been performed for the data rebinned to a $16^3$ grid. The results are
presented in Table 1. 

The average temperature for the rebinned data seems to have converged already in
the $64^3$ simulation. However, the behaviour of the rms density fluctuation
suggests that a higher resolution is needed for the convergence of the density
field. The variation of the density weighted temperatures between the $64^3$
and $128^3$ grids depends essentially on variations of density distribution. As
already noticed by Kang et al. (1994) for all the eulerian codes considered in
that paper, the values of $\sigma$ appear to converge from  below. 

In Fig. 5 we show a cell--by--cell comparison of density and temperature for
the rebinned data of the three simulations. We always observe a good
correlation between the values of the densities obtained from simulations with
different resolutions; the spread of the points is reduced as we compare
simulations with the highest resolution. The same holds for the high
temperature cells. The distribution at lower temperatures is due to the
formation of substructures as we increase the grid resolution. The large number
of points which accumulate in the lower part of the panel represents cells that
in the lower resolution simulation are cooling according to the expansion of
the Universe while in the higher resolution run are site of some structure.
Also this effect decreases when the resolution is higher. 

In the left and right columns of Fig. 6 we show the volume and mass weighted
density histograms, $f(N)$ and $f(M)$ respectively, 
for the $16^3$ rebinned data. The tendency to
form higher density peaks as we increase the resolution clearly appears in
these histograms. As we go from $32^3$ to $128^3$ the amount of mass contained
in the high density regions increases. However, in relation to the convergence
of the numerical results, it is encouraging to see that the density
distribution both in volume and in mass is very similar 
for the $64^3$ and $128^3$ runs.

The corresponding histograms for the temperature are shown in Fig. 7. Comparing
the rebinned data from different runs we can observe higher gas temperatures
when the resolution of the simulation is increased.
We notice both an increase in
volume fraction and in mass fraction of the gas at high temperature. The $f(N)$
peak at low temperature is almost absent in the $128^3$ simulation, but this is
only due to the averaging procedure in constructing the rebinned data. Small
scale structures are present in the high resolution run which could not appear
in the simulations with a smaller number of cells and they have the effect of
raising the average temperature in the rebinned data. 

The same effect is evident if we look at Fig. 8 which presents the same
histograms for the original data of the $128^3$ simulation: a large fraction of
the volume is still at low temperature. However, the mass fraction distribution
confirms that about $70 \%$ of the mass is at temperatures higher than $10^6$
K. 

In Fig. 9 we show two different slices of the simulation with $0.5~h^{-1}$ Mpc
thickness (1 cell). For each slice contour plots are presented for 
the density contrast $\delta=\varrho/\langle
\varrho\rangle -1$ and the temperature. Structures
appear well resolved and strong temperature gradients indicate shock
positions. High density peaks are surrounded by high temperature regions which
have been heated by the shock propagation. 
 
Finally in Fig. 10 the contour plots of the volume (left) and mass (right)
fraction with given
temperature and density are shown. We observe that most of gas is at high
density and temperature. On the other hand most of the volume is at low density
with either high or low temperatures depending on whether it has been 
crossed by
shocks or not.

\section{Conclusions}

We have presented a new numerical code developed for studying the 
formation and evolution of cosmological structures in both baryonic and 
dark
components. Collisional matter is treated as a fluid and the corresponding
hydrodynamic equations are solved using the PPM scheme on a fixed 
Eulerian grid.
We have described the changes to the basic method required by
the cosmological applications. Particular care has been taken in including
expansion and gravity in the Riemann solver and in the final integration step.
This has required the calculation of the characteristic form of the
hydrodynamic equations in expanding coordinates. 
A double formulation of the energy equation has allowed a proper 
treatment of the highly supersonic flows common in cosmological simulations.
The behaviour of the dark matter is described using a standard Particle Mesh
N--body technique, modified to allow the use of a variable timestep,
as desirable for hydrodynamics.
The two components are coupled through the
gravitational interaction and the gravitational field is calculated 
from the Poisson equation using an FFT procedure. 

We have presented a series of tests selected for their relevance in
cosmological applications, paying attention both to the accuracy of the highest
resolution results and to the convergence of the method when lower resolutions
are used. 

The one--dimensional tests show that the code can reproduce properly the
expected solutions, even when very low resolution is adopted. In particular we
present the results of the shock tube test and of single and multiple
pancake formation. The CDM test results can be compared with those presented
by Ryu et al. (1994), Kang et al. (1994) and Gnedin (1995) showing a 
good agreement. All of these tests have demonstrated that our code can be 
considered a reliable and useful tool for cosmological studies.

\section* {Acknowledgments} We would like to thank John Miller for 
many helpful discussions and comments. 
This work was partially supported by Italian
MURST.

\bigskip

\section*{Figure captions}
\noindent

\noindent
{\bf Figure 1.} The velocity $v$, the pressure $p$ and the density $\varrho$
for the one--dimensional shock tube test at the final time $t_0=20$. 
The numerical result obtained with a grid of 128 zones (open circles) is 
compared to the analytical solution (solid line). 

\noindent
{\bf Figure 2.} The velocity $v$, the temperature $T$ and the density $\varrho$
for the one--dimensional Zel'dovich pancake test at the final time $a_0=1$. The
results obtained with grids of 256 (open circles) and 1024 zones (solid line)
are compared. 

\noindent
{\bf Figure 3.} The energy conservation parameter $R$ (top left), 
the gravitational energy
$W$ (top right), the kinetic energy $E_{KIN}$ (down left) and the 
thermal energy $E_{TH}$ (down right) for the
one--dimensional Zel'dovich pancake test are plotted 
as function of the number of cells $N_c$. 

\noindent
{\bf Figure 4.} The velocity $v$, the temperature $T$ and the density $\varrho$
for the one--dimensional multiple pancake test at the final time $a_0=1$. A
primordial power--law power spectrum with spectral index $n=1$ is simulated
using a grid of 256 zones. 

\noindent
{\bf Figure 5.} A cell--by--cell comparison of density $\varrho$ (upper row)
and temperature $T$ (lower row) for the $16^3$ rebinned data at $z=0$ for 
three simulations with the same initial conditions but different numbers of
zones, as indicated by the subscripts. 

\noindent
{\bf Figure 6.} The volume--weighted $f(N)$ (left column) and mass--weighted
$f(M)$ (right column) histograms at $z=0$ for the density $\varrho$ 
at $z=0$ for the $16^3$
rebinned data. The results obtained with different numbers of zones are shown:
$N_c=32$ (upper row), $N_c=64$ (central row) and $N_c=128$ (lower row). 

\noindent
{\bf Figure 7.} The same as Fig. 6 but for the temperature $T$. 

\noindent
{\bf Figure 8.} The volume--weighted $f(N)$ (upper panels) and mass--weighted
$f(M)$ histograms (lower panels) for density $\varrho$ (left column) and
temperature $T$ (right column) at $z=0$ for the original unbinned
data of the $128^3$ simulation. 

\noindent
{\bf Figure 9.} Density (left column) and temperature (right column) 
contour plots of two different slices
of $0.5 \ h^{-1}$ Mpc thickness in the $128^3$ run are shown.
Density is normalized to the mean density while
the temperature is in units of $10^6$ K. The density contour levels 
correspond to $10^{(i-3)/4}$, while the temperature levels are 
$5^{i/2}$ (solid lines) and $5^{-2i}$ (dashed lines), where $i=1,2,...$

\noindent
{\bf Figure 10.} Contour plots of the volume fraction (left) 
and mass fraction (right)
with given temperature and density. The results at $z=0$ from the $128^3$
simulation are shown. Contour levels correspond to $10^{i/4}$, where  
$i=1,2,...$. 

\bigskip
\appendix

\section{Numerical scheme}

We here 
describe the main steps in the integration of the dynamical equations presented
in the text. We used a modified PPM scheme to follow the evolution of baryonic
matter and a PM N--body method for the dark matter. The same grid of constant
mesh size $\Delta x$ is used for solving the hydrodynamical equations and
for 
computing the total gravitational potential. The subscript $j$ indicates zone
centred values of grid quantities, while subscript $(j+1/2)$ refers to values
computed at the boundary between the $j$--th and $(j+1)$--th zones. 
Finally, in the dynamical
equations for the dark matter the subscript $j$ refers to the $j$--th particle.
Superscripts indicate the time level at which variables are calculated. The
same time interval $\Delta t^{n+1/2}=t^{n+1}-t^n$ is used for integrating the
dynamical equations of both components. The time--integration of the N--body
dynamical equations is performed by a Lax--Wendroff scheme instead of the usual
leapfrog scheme which would require a constant timestep. 

The main steps in the integration of the dynamical equations are the 
following:
\smallskip

\noindent {\bf 1} -- 
At the time $t^n$ we have $\varrho^n_{BM,j}$, ${\bf v}^n_{BM,j}$ and
${p^n_j}$ for the baryonic matter.  For the dark matter we have ${\bf
x}^n_{DM,j}$ and ${\bf v}^n_{DM,j}$. First we calculate the dark
matter density $\varrho^n_{DM,j}$ using a cloud--in--cell mass
assignment scheme. Then we compute the total peculiar gravitational
potential $\phi^n_j$ on the grid by solving the Poisson equation using 
a standard FFT technique. We also
extrapolate the gravitational potential to $t^{n+1/2}$ using the values
at $t^n$ and $t^{n-1}$ 
\be
\phi^{n+1/2}\ =\ \phi^{n}\left (1+{\Delta t^{n+1/2}\over 2\Delta t^{n-1/2}}
\right ) - \phi^{n-1}{\Delta t^{n+1/2}\over 2\Delta t^{n-1}}\ .%\eqno(G1)
\ee

\smallskip
\noindent {\bf 2} -- We start the hydrodynamical one--dimensional 
sweep. For a given
direction we indicate by $v$ the component of the velocity along the
direction of integration and by $u$ the component orthogonal to $v$. We then
calculate the time averaged values of the hydrodynamical quantities at the zone
interfaces $\bar\varrho_{j+1/2}$, $\bar p_{j+1/2}$, $\bar v_{j+1/2}$ and $\bar
u_{j+1/2}$ using the PPM scheme, as described in Section 3.1, 3.2 and in
Appendix A2.

\smallskip
\noindent {\bf 3} -- We use the time averaged estimates calculated in point 2 to
solve the one--dimensional hydrodynamical equations without the inclusion 
of the expansion terms (for simplicity in the following equations 
we have dropped the $BM$ subscript): 
\be
\varrho_{H,j}^{n+1}=\varrho_{j}^{n}+ {\Delta t^{n+1/2}\over a^{n+1/2}}
\left ({ \brhon_{j-1/2}\bv_{j-1/2}-\brhon_{j+1/2}\bv_{j+1/2}\over\Delta x }
\right )
\ee
\be
v_{H,j}^{n+1}={1\over \varrho_{H,j}^{n+1}} 
\left [\varrho_{j}^{n} v_j^n+{\Delta t^{n+1/2}\over a^{n+1/2}}\left (
{ \brhon_{j-1/2}\bv_{j-1/2}^2+\bp_{j-1/2}-\brhon_{j+1/2}\bv_{j+1/2}^2
-\bp_{j+1/2}\over\Delta x }\right )+{\varrho_{j}^{n+1}+\varrho_{j}^{n}
\over 2}g_j \right ]
\ee
\be
E_{H,j}^{n+1}=E_{j}^{n}+{\Delta t^{n+1/2}\over a^{n+1/2}}\left [
{ \left({1\over 2}\brhon_{j-1/2}\bv_{j-1/2}^2+{\gamma\over \gamma -1}\bp_{j-1/2}
\right)\bv_{j-1/2}-\left ({1\over 2}\brhon_{j+1/2}\bv_{j+1/2}^2+
{\gamma\over \gamma -1}\bp_{j+1/2}\right)\bv_{j+1/2}
\over\Delta x }\right ]
 +{{\varrho_{j}^{n+1}v_{j}^{n+1}+\varrho_{j}^{n}
v_{j}^{n}}\over 2}g_j
\ee
\be
u_{H,j}^{n+1}\ =\ {1\over \varrho_{H,j}^{n+1}} 
\left [\varrho_{j}^{n} u_j^n+{\Delta t^{n+1/2}\over a^{n+1/2}}\left (
{ \brhon_{j-1/2}\bv_{j-1/2}\bu_{j-1/2}-\brhon_{j+1/2}\bv_{j+1/2}\bu_{j+1/2}
\over\Delta x }\right )\right ]\ .%\eqno(P2)
\ee
The quantity $g_j$ is the time--centred value
of the gravitational force averaged on the $j$--th cell calculated
by expanding the density and acceleration fields and retaining all terms up to
second--order in $\Delta x$:
\be
g_j\ =\ {1\over 2a(t)\Delta x}\left [ \phi_{j+1}-\phi_{j-1}+{1\over 12}
(\phi_{j+1}-2\phi_{j}+\phi_{j-1}){\delta\varrho_j\over\varrho_j}\right ]\ .
%\eqno(G2)
\ee
The monotonicity of the numerical method is preserved by computing the density
slope $\delta\varrho$ as:

\begin{eqnarray}
\delta\varrho_j \  =&{\tilde 
\varrho_{min}} {\rm sign} (\varrho_{j+1}-\varrho_{j-1}) 
&{\rm if}(\varrho_{j+1}-\varrho_{j})(\varrho_{j}-\varrho_{j-1}) < 0
\nonumber \\
  =&0 & \  {\rm otherwise}\ ,
%\eqno(G3)
\end{eqnarray}
where ${\tilde \varrho_{min}}= \min(\vert \varrho_{j+1}-\varrho_{j-1}\vert,
2\vert\varrho_{j}-\varrho_{j-1}\vert,2\vert\varrho_{j+1}-\varrho_{j}\vert)$.
Because of its smoothness, the gravitational potential does not require the same
monotonicity constraint.
 
\smallskip
Points 2 and 3 are repeated for the other two orthogonal directions completing
the three--dimensional hydrodynamical calculation.

\smallskip
\noindent {\bf 4} -- The hydrodynamical quantities are corrected by
considering the expansion terms:
\be
\varrho_j^{n+1} \ =\ \varrho_{H,j}^{n+1}
\ee
\be
v_{j}^{n+1} \ =\ {2v_{H,j}^{n+1}-\Delta t^{n+1/2}(\dot a/a)^{n+1/2}v_j^{n}\over
2+\Delta t^{n+1/2}(\dot a/a)^{n+1/2}}
\ee
\be
E_{j}^{n+1} \ =\ {E_{H,j}^{n+1}-\Delta t^{n+1/2}(\dot a/a)^{n+1/2}E_j^{n}\over
1+\Delta t^{n+1/2}(\dot a/a)^{n+1/2}}
%\eqno(A0)
\ee

\smallskip
\noindent {\bf 5} -- Position and velocity of the 
N--body particles are updated by 
the two--step Lax--Wendroff scheme:

\noindent step 1
\be
{\bf v}_{DM,j}^{n+1/2}\ =\ {\bf v}_{DM,j}^{n}- \left(
\dot a^n {\bf v}_{DM,j}^{n}+{\bf g}_{j}^n \right)
{\Delta t^{n+1/2}\over 2 a^n}
\ee
\be
{\bf x}_{DM,j}^{n+1/2}\ =\ {\bf x}_{DM,j}^{n}+{\bf v}_{DM,j}^{n}
{\Delta t^{n+1/2}\over 2 a^n}
\ee
step 2
\be
{\bf v}_{DM,j}^{n+1}\ =\ {\bf v}_{DM,j}^{n}- \left( 
\dot a^{n+1/2} {\bf v}_{DM,j}^{n+1/2}+{\bf g}_{j}^{n+1/2}\right)
{\Delta t^{n+1/2}\over a^{n+1/2}}
\ee
\be
{\bf x}_{DM,j}^{n+1}\ =\ {\bf x}_{DM,j}^{n}+
{\bf v}_{DM,j}^{n+1/2}{\Delta t^{n+1/2}\over a^{n+1/2}}
\ee
Here we indicate by ${\bf g}_{j}^n$ the gravitational
force on the $j$--th particle, computed by interpolating the values
of the corresponding components at the eight neighbouring cells 
using the cloud--in--cell scheme.

\smallskip
\noindent {\bf 6} -- Energy is corrected to account for highly supersonic flows.
We have introduced the internal energy equation.
This is solved first including the flux term by the
standard PPM integration procedure:
\be
\tilde e_{H,j}^{n+1}=e_{j}^{n}+{\Delta t^{n+1/2}\over a(t^{n+1/2})}\left (
{ \bp_{j-1/2}\bv_{j-1/2}-\bp_{j+1/2}\bv_{j+1/2}
\over(\gamma -1)\Delta x }\right )\ .
\ee
Then this value is corrected as:
\be
e^{n+1}_{H,j}\ =\tilde e^{n+1}_{H,j}+{1\over a^{n+1/2}}{\Delta t^{n+1/2}
\over\Delta x}
{p_j^{n+1/2}\over\varrho^n_j}(\bar v_{j-1/2}-\bar v_{j+1/2})\ ,
\ee 
where $p_j^{n+1/2} = (\bp_{j-1/2}+\bp_{j+1/2})/2$.
Finally, we consider also the expansion term in the internal energy equation:
\be
e_{j}^{n+1} \ =\ {e_{H,j}^{n+1}-\Delta t^{n+1/2}(\dot a/a)^{n+1/2}e_j^{n}\over
1+\Delta t^{n+1/2}(\dot a/a)^{n+1/2}}\ .
\ee
The internal energy is used if conditions (27) hold:
\smallskip

\be
\sum_{i=1,3}{\partial v_j\over\partial x_j}\ \ge\ 0\ ;
\qquad{1\over p_j}\left [\sum_{i=1,3}\left ({\partial p_j\over \partial x_i})^2
\right )\right ]^{1/2}\ \le\ {\eta_1}\qquad $j=1,3$.
\ee
If neither of these conditions are verified, internal 
energy is computed according to expression (28):
\be
e_j =\cases{
E_j-{1\over 2}\varrho_j{\bf v}^2\quad 
& if $E_j-{1\over 2}\varrho_j{\bf v}^2/ \max( E_{j-1},E_{j},E_{j+1})
\geq \eta_2$\cr
e_j & otherwise\ .\cr}%\eqno(E5)
\ee
Note that as we
are dealing with errors that can be advected with the total energy calculation,
we have to consider not just the $j$--th cell but also its neighbourhood (that
in one dimension comprises the two zones $j\pm 1$). 

\section{Calculating Fluxes}

Here we describe briefly the procedure used to calculate the time
averaged values of the hydrodynamical quantities at the zone
interfaces $\bar\varrho_{j+1/2}$, $\bar p_{j+1/2}$, $\bar v_{j+1/2}$ and
$\bar u_{j+1/2}$ . We include to second--order 
both the gravity and the cosmic expansion, which
enter the characteristic equations as described in Section 3.2.

We indicate by $w(x)$ the interpolated piecewise continuous function
representing the generic hydrodynamical variable. The quantity
$w_j$ represents the mean value of $w(x)$ over the $j$--th zone.

We define $\tilde w_{j+1/2,L}$ and $\tilde w_{j+1/2,R}$, first
approximations to the effective left and right states of the Riemann
problem, as:
\be
\tilde w_{j+1/2,L}\ =\  f_{j+1/2,L}(x_{j+1/2}-\tilde x_{j+1/2,L})\ ,
\ee
\be
\tilde w_{j+1/2,R}\ =\  f_{j+1/2,R}(-x_{j+1/2}+\tilde x_{j+1/2,R})\ ,
\ee
where 
\be
f_{j+1/2,L}(\Delta x)={1\over 
\Delta x}\int_{x_{j+1/2}-\Delta x}^{x_{j+1/2}}w(\xi)~d\xi,\qquad\qquad
f_{j+1/2,R}(\Delta x)
={1\over \Delta x}\int^{x_{j+1/2}+\Delta x}_{x_{j+1/2}}w(\xi)~d\xi\ ,
\ee
\be
\tilde x_{j+1/2,L}\ =\ x_{j+1/2}-\max(0,\Delta t^{n+1/2}~\lambda_{j,+})\ ,\qquad
\tilde x_{j+1/2,R}\ =\ x_{j+1/2}+\max(0,-\Delta t^{n+1/2}~\lambda_{j+1,-})
\ee
and
\be
\lambda_{j,+}\ =\ {v^n_j+c_j^n\over a^n}\ , \qquad
\lambda_{j+1,-}\ =\ {v^n_{j+1}-c_{j+1}^n\over a^n}\ .
\ee
We correct our initial guess for the left and right states by solving
the equations of gas dynamics in characteristic form (see Section 3.2).
First we calculate the mean value of the hydrodynamic variables 
in the domains defined by each of the characteristic lines coming
from the left and the right of the zone boundary $j+1/2$:
\be
w_{j+1/2,L}^k\ =\  f_{j+1/2,L}(x_{j+1/2}-x_{j+1/2,L}^k)\ ,
\ee
\be
w_{j+1/2,R}^k\ =\  f_{j+1/2,R}(-x_{j+1/2}+x_{j+1/2,R}^k)\ ,
\ee
where $k=+,0,-$ and
\be
x_{j+1/2,L}^k\ =\ x_{j+1/2}-\Delta t\lambda_{j,k},\qquad
x_{j+1/2,R}^k\ =\ x_{j+1/2}-\Delta t\lambda_{j+1,k}\ .
\ee
Then we correct the initial guess considering only that information
which can reach the zone edge during the current timestep:
\be
p_{j+1/2,S}\ =\ \tilde p_{j+1/2,S}+\tilde C_{j+1/2,S}^2(
\beta_{j+1/2,S}^+ +\beta_{j+1/2,S}^-)\ ,
\ee   
\be
v_{j+1/2,S}\ =\ \tilde v_{j+1/2,S}+\tilde C_{j+1/2,S}(
\beta_{j+1/2,S}^+ -\beta_{j+1/2,S}^-)\ ,
\ee   
\be
\varrho_{j+1/2,S}\ =\ \left ({1\over\tilde \varrho_{j+1/2,S}}-\sum_{k=+,0,-}
\beta_{j+1/2,S}^k\right )^{-1}\ .
\ee
Here $\tilde C_{j+1/2,S}^2=\gamma\tilde p_{j+1/2}\tilde \varrho_{j+1/2}$ and
$S=L,R$. We also have
\be
u_{j+1/2,L}\ =\ \tilde u_{j+1/2,L},\qquad \beta_{j+1/2,L}^k=0\quad
{\rm if}\quad \lambda_{j,k}\le 0\ ,
\ee
\be
u_{j+1/2,R}\ =\ \tilde u_{j+1/2,R},\qquad \beta_{j+1/2,R}^k=0\quad
{\rm if}\quad \lambda_{j+1,k}\le 0\ .
\ee
Otherwise
\be
\beta_{j+1/2,S}^{\pm}\ =\ \mp {1\over \tilde C_{j+1/2,S}}\left [
(\tilde v_{j+1/2,S}-v_{j+1/2,S}^{\pm})\pm {\tilde p_{j+1/2,S}-p_{j+1/2,S}
^{\pm}\over\tilde C_{j+1/2,S}}\pm \Delta t\left(2{\dot a\over a}
{p_{j+1/2,S}^{\pm}\over \tilde C_{j+1/2,S}}\pm {\dot a\over a} 
v_{j+1/2,S}^{\pm}\mp g_{j+1/2,S}^{\pm}\right )\right ]
\ee
\be
\beta_{j+1/2,S}^{0}\ =\ \left({\tilde p_{j+1/2,S}-p_{j+1/2,S}^0\over
\tilde C_{j+1/2,S}^2}+{1\over \tilde\varrho_{j+1/2,S}}-
{1\over\varrho_{j+1/2,S}^0}+2{\dot a\over a}{p_{j+1/2,S}^{\pm}\over 
\tilde C_{j+1/2,S}^2}\right )
\ee
\be
u_{j+1/2,S}\ =\ u_{j+1/2,S}^0\ .
\ee
The time averaged estimates of the solution at $x_{j+1/2}$ are determined
by 
calculating the non--linear interaction of the left and right states calculated
in equations (A33)--(A35). This interaction is described by the solution to the
Riemann problem with these left and right states
as initial conditions. 
The Riemann problem is solved in a standard way, as described by 
Van Leer (1979) and Gottlieb \& Groth (1988).

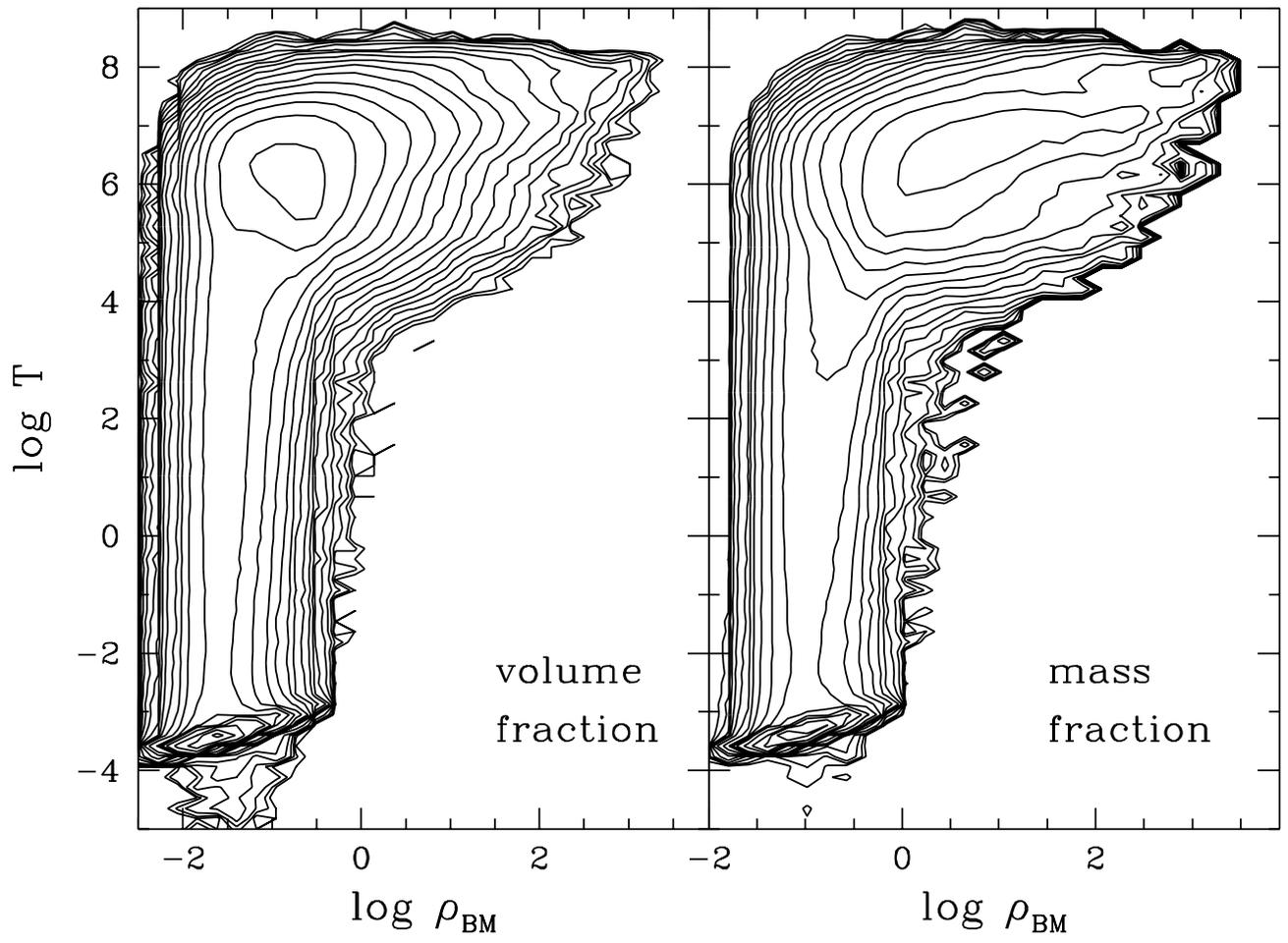
\end{document}